\documentclass[%
 reprint,
superscriptaddress,
nofootinbib,
 amsmath,amssymb,
 aps,
prb,
]{revtex4-2}

\usepackage{graphicx}
\usepackage{dcolumn}
\usepackage{bm}

\usepackage{hyperref}
\usepackage{orcidlink}
\usepackage[T1]{fontenc}
\usepackage{listings}
\usepackage{colortbl}
\usepackage{float}
\usepackage{xcolor}

\hypersetup{colorlinks=true,citecolor=green,linkcolor=red,urlcolor=blue}

\newcommand{\vela}{\texttt{Vela.jl}}
\newcommand{\pint}{\texttt{PINT}}
\newcommand{\pyvela}{\texttt{pyvela}}

\newcommand{\tempo}{\texttt{tempo}}
\newcommand{\tempotwo}{\texttt{tempo2}}
\newcommand{\temponest}{\texttt{TEMPONEST}}

\newcommand{\emcee}{\texttt{emcee}}
\newcommand{\astropy}{\texttt{astropy}}

\newcommand{\updated}[1]{
#1
}

\definecolor{codegreen}{rgb}{0,0.6,0}
\definecolor{codegray}{rgb}{0.5,0.5,0.5}
\definecolor{codepurple}{rgb}{0.58,0,0.82}
\definecolor{backcolour}{rgb}{0.95,0.95,0.92}
\lstdefinestyle{pystyle}{
  backgroundcolor=\color{backcolour}, commentstyle=\color{codegreen},
  keywordstyle=\color{magenta},
  numberstyle=\tiny\color{codegray},
  stringstyle=\color{codepurple},
  basicstyle=\ttfamily\footnotesize,
  breakatwhitespace=false,         
  breaklines=true,                 
  captionpos=b,                    
  keepspaces=true,                 
  numbers=left,                    
  numbersep=5pt,                  
  showspaces=false,                
  showstringspaces=false,
  showtabs=false,                  
  tabsize=2
}
\newfloat{lstfloat}{}{lop}
\floatname{lstfloat}{Code}

\begin{document}

\preprint{APS/123-QED}

\title[Bayesian wideband timing with \vela{}]{Bayesian pulsar timing and noise analysis with \vela{}: the wideband paradigm}

\author{Abhimanyu Susobhanan\orcidlink{0000-0002-2820-0931}}
\affiliation{Max-Planck-Institut f{\"u}r Gravitationsphysik (Albert-Einstein-Institut), Leibniz Universit{\"a}t Hannover, Callinstra{\ss}e 38, 30167 Hannover, Deutschland}

\date{\today}

\begin{abstract}
\vela{} is a package for performing Bayesian pulsar timing and noise analysis written in Julia and Python. 
In the wideband paradigm of pulsar timing, simultaneous time of arrival and dispersion measure measurements are derived from a radio observation using frequency-resolved integrated pulse profiles and templates without splitting the observation into multiple frequency sub-bands.
We describe the implementation of the wideband timing paradigm in \vela{}, and demonstrate its usage using the NANOGrav 12.5-year wideband data of PSR J1923+2515.
\vela{} is the first software package to provide this functionality.
\end{abstract}

\maketitle


\section{Introduction}
\label{sec:intro}
Pulsars are rotating neutron stars whose broad-band electromagnetic radiation is received as periodic pulses by terrestrial observers \citep{LorimerKramer2012}.
In addition to being fascinating objects in their own right with rich phenomenology, their high rotational stability allows us to use them as celestial clocks to study a wide range of time-domain astrophysical effects ranging from solar wind variability \citep[e.g.][]{SusarlaChalumeau+2024} to gravitational waves \citep[e.g.][]{AgazieAntoniadis+2024}.
This is possible through pulsar timing,  the technique of accurately tracking the rotation of a pulsar by measuring the times of arrival (TOAs) of their pulses \citep{HobbsEdwardsManchester2006}.

In addition to the pulsar rotation, several astrophysical and instrumental effects influence the measured TOAs, including the binary motion of the pulsar, interstellar dispersion, solar system motion, the observatory clock, instrumental noise, etc \citep{EdwardsHobbsManchester2006}.
Interstellar dispersion, for example, introduces a delay $\Delta_{\texttt{DM}}=\mathcal{K}\mathcal{D}/\nu^2$ to the TOA, where $\mathcal{K}$ is known as the dispersion constant, $\mathcal{D}$ is the electron column density along the line of sight known as the dispersion measure (DM), and $\nu$ is the observing frequency in the solar system barycenter (SSB) frame.
Pulsar timing (or more generally, pulsar timing and noise analysis) involves accurately modeling all of these contributions, and a mathematical description of these effects is called a pulsar timing and noise model.

In conventional pulsar timing, also known as narrowband timing, the fundamental measurable quantity is the narrowband TOA.
A frequency-resolved radio lightcurve obtained from a pulsar observation is folded using the known pulse period to obtain the integrated pulse profile, which gives the pulsar intensity as a function of observing frequency and pulse phase \citep{HotanVan_StratenManchester2004}.
Narrowband TOAs are measured by dividing an integrated pulse profile into multiple frequency sub-bands and cross-correlating the profile against a noise-free template in each sub-band \citep{Taylor1992}.
Template profiles are usually derived from data, and both frequency-resolved \citep[e.g.][]{TarafdarNobleson+2022} and frequency-averaged \citep[e.g.][]{DemorestFerdman+2013} templates have been used in the literature.
See, e.g., \citet{WangShaifullah+2022} for comparisons of different narrowband TOA generation methods.

Recent years have seen significant improvements in telescope sensitivity both through upgrades to existing observatories and the commissioning of new ones, including the advent of wide-band receivers and backends, reduction in the system noise, and large collecting areas \citep[e.g., ][]{HobbsManchester+2020,LiuJiang+2022}.
Such improvements, along with long-running pulsar timing experiments like Pulsar Timing Arrays \citep[PTAs:][]{FosterBacker1990,PereraDeCesar+2019}, have posed significant challenges to the narrowband timing paradigm.
This includes inadequate modeling of frequency-dependent profile evolution \citep{HankinsRickett1986} and interstellar scattering \citep{HembergerStinebring2008}, as well as large data volumes \citep[e.g.][]{AgazieAlam+2023}.
The wideband timing method addresses some of these challenges by treating the integrated pulse profile as a single two-dimensional entity rather than as a collection of independent sub-bands \citep{PennucciDemorestRansom2014,LiuDesvignes+2014}.
In this paradigm, a single TOA and a single DM are measured from an integrated pulse profile using the full bandwidth of the observation, and the wideband DM measurement is treated as an additional data point together with the wideband TOA.
This can reduce the data volume by up to two orders of magnitude compared to the narrowband paradigm since TOAs are no longer measured per sub-band. 
Wideband TOAs and DMs are measured with the help of two-dimensional templates usually derived from the data using principal component analysis \citep{Pennucci2019}.
This method has been shown to significantly improve the representation of the frequency-dependent profile evolution in the template.
It is also possible to derive a wideband TOA-DM pair by combining simultaneous multi-band observations \citep{PaladiDwivedi+2023}.
It should be noted that pulse broadening due to interstellar scattering \citep[e.g.][]{KrishnakumarMitra+2015} is not yet adequately handled in the wideband paradigm\footnote{\updated{Rigorous treatment of interstellar scattering is an open problem in pulsar timing, and it is only modeled approximately even in the narrowband paradigm \citep{HembergerStinebring2008}. Promising avenues for handling interstellar scattering under active investigation include cyclic spectroscopy \citep[e.g.][]{TurnerStinebring+2023} and CLEAN algorithm-based deconvolution \citep[e.g.][]{YoungLam2024}.}}.

In practice, the wideband TOAs and DMs are usually measured using the \texttt{PulsePortraiture} package \citep{Pennucci2019,PennucciDemorestRansom2014}.
Wideband timing methods, described in Section \ref{sec:wb-timing}, are available in the \tempo{} \citep{NiceDemorest+2015} and \pint{} \citep{LuoRansom+2021,SusobhananKaplan+2024} packages, whereas the \tempotwo{}  package \citep{HobbsEdwardsManchester2006,EdwardsHobbsManchester2006} does not support wideband timing.
Bayesian noise analysis for wideband datasets, using a linear approximation of the timing model, is available in the \texttt{ENTERPRISE} package \citep{EllisVallisneri+2020,JohnsonMeyers+2024}.
The \temponest{} package \citep{LentatiAlexander+2014}, which provides Bayesian timing and noise analysis functionality for narrowband datasets without the linear approximation, does not support the wideband paradigm.
Further, all the wideband timing and noise analysis efforts so far \citep[e.g.][]{AlamArzoumanian+2021,TarafdarNobleson+2022,CuryloPennucci+2023} have exclusively used a piecewise-constant model \citep[DMX:][]{ArzoumanianBrazier+2015} to describe the DM variations.
Gaussian process models of DM variability are not yet supported in \texttt{ENTERPRISE}, although a method for their implementation was recently presented in \citet{SusobhananVan_Haasteren2025}.

\vela{}\footnote{The source code is available at \url{https://github.com/abhisrkckl/Vela.jl}. The documentation is available at \url{https://abhisrkckl.github.io/Vela.jl/}.} is a new Bayesian pulsar timing and noise analysis package written in Julia and Python, first presented in \citet{Susobhanan2025}.
In this paper, we present, for the very first time, a public implementation of the full non-linear wideband timing and noise model geared towards Bayesian inference in \vela{}.
We aim to provide a fully general implementation of the wideband timing and noise model that can support arbitrary DM models, including interstellar and solar wind dispersion models, both deterministic and stochastic, \updated{incorporating the method developed in \citet{SusobhananVan_Haasteren2025}}.
We also summarize the new developments in \vela{} after the publication of \citet{Susobhanan2025} corresponding to both narrowband and wideband paradigms.

This paper is arranged as follows. 
Section \ref{sec:wb-timing} provides a brief overview of wideband timing and noise analysis, and Section \ref{sec:vela-impl} describes its implementation in \vela{}.
We demonstrate this implementation using a real dataset in Section \ref{sec:sims}, and we summarize our work in Section \ref{sec:summary}.
Appendix \ref{sec:pyvela-cli} describes a command-line interface for \vela{}.
A glossary of symbols used in this paper is provided as supplementary material.

\section{A Brief Overview of Bayesian Wideband timing and noise Analysis}
\label{sec:wb-timing}

A wideband measurement includes the measured TOA and DM values, the corresponding uncertainties, the observing frequency, and metadata about the observing system stored as flags.
It is important to note that a wideband TOA without its DM is not the same as a narrowband TOA and does not contain enough information to do timing accurately.
Wideband TOAs are usually stored and distributed as \texttt{tim} files in a format that is an extension of the narrowband TOA format.

A wideband TOA $t_{\text{arr}}$ measured at a terrestrial observatory is related to the pulse emission time $t_{\text{em}}$ as
\begin{align}
    t_{\text{arr}} &= t_{\text{em}} + \Delta_{\text{B}}  + \Delta_{\text{DM}}  + \Delta_{\text{GW}} + \Delta_{\odot} + \Delta_{\text{clock}}  \nonumber\\
    &\quad + \Delta_{\text{jump}} + \mathcal{N}_{\text{R}} +  ... \,,
    \label{eq:delays}
\end{align}
where $\Delta_{\text{B}}$ contains delays arising from the binary motion of the pulsar \citep{DamourDeruelle1986}, 
$\Delta_{\text{DM}}$ is the total dispersion delay including interstellar dispersion and solar wind dispersion \citep{BackerHellings1986},
$\Delta_{\text{GW}}$ represents the delays caused by gravitational waves propagating across the line of sight to the pulsar \citep{EstabrookWahlquist1975},
$\Delta_{\odot}$ contains the delays arising from solar system motion \citep{EdwardsHobbsManchester2006},
$\Delta_{\text{clock}}$ is a clock correction that converts the TOA measured against an observatory clock into an international time standard \citep{HobbsEdwardsManchester2006}, and $\Delta_{\text{jump}}$ contains the various instrumental delays.
$\mathcal{N}_{\text{R}}$ is a time-uncorrelated stochastic term that represents the radiometer noise \citep{LorimerKramer2012}.
Note that we have ignored an unmeasurable constant term representing the vacuum light travel time from the pulsar to the observatory at a fiducial epoch in the above equation.

The wideband DM measurement $d$ is given by
\begin{equation}
    d = \mathcal{D} + \mathcal{D}'_{\text{jump}} + \mathcal{M}_{\text{R}} + \mathcal{M}_{\text{jitter}}\,,
    \label{eq:dmmodel}
\end{equation}
where $\mathcal{D}$ is the total astrophysical DM, $\mathcal{D}'_{\text{jump}}$ represents the system-dependent wideband DM offsets \citep{AlamArzoumanian+2021}, and $\mathcal{M}_{\text{R}}$ and $\mathcal{M}_{\text{jitter}}$ are time-uncorrelated stochastic terms arising due to radiometer noise and pulse jitter noise (random pulse shape variations) \citep{ParthasarathyBailes+2021}.
$\mathcal{D}$ is given by
\begin{equation}
    \mathcal{D} = \sum_{n=0}^{n_D} \frac{1}{n!} D_n (t_{\text{em}}-t_0)^{n} + D_{\text{DMN}} + D_{\text{SW}}+ ...\, 
\end{equation}
where $D_n$ are Taylor coefficients that determine the long-term evolution of the astrophysical DM, $D_{\text{DMN}}$ represents the stochastic DM variations (DM noise), and $D_{\text{SW}}$ represents the solar wind DM contribution.
$\mathcal{D}$ is related to $\Delta_{\text{DM}}$ as $\Delta_{\text{DM}}=\mathcal{K}\mathcal{D}/\nu^2$ as mentioned in the introduction.

The rotational phase $\phi$ can be written as \citep{HobbsEdwardsManchester2006}
\begin{align}
    \phi &= \phi_0 + \sum_{n=0}^{n_F} \frac{1}{(n+1)!} F_n (t_{\text{em}}-t_0)^{n+1} + \phi_{\text{glitch}} \nonumber\\
    &\quad + \phi_{\text{SN}} + \mathcal{N}_{\text{jitter}} + ...\,
    \label{eq:phasing}
\end{align}
where $\phi=1$ represents a full rotation, $\phi_0$ is an initial phase, $F_n$ represent the rotational frequency and its derivatives, $\phi_{\text{glitch}}$ represents a phase correction due to glitches, $\phi_{\text{SN}}$ represents the slow stochastic wandering of the rotational phase known as spin noise, and $\mathcal{N}_{\text{jitter}}$ is an uncorrelated noise term due to pulse jitter.
The timing residual $r$ and the DM residual $\delta$ can be written as
\begin{subequations}
\begin{align}
    r &= \frac{\phi - \mathcal{N}[\phi]}{\bar{F}}\,,
    \label{eq:resids}\\
    \delta &= d - \mathcal{D} - \mathcal{D}'_{\text{jump}}\,,  \label{eq:dm-resids}  
\end{align}
\end{subequations}
where $\mathcal{N}[\phi]$ is the closest integer to $\phi$ and $\bar{F}=d\phi/dt_{\text{arr}}$ is the topocentric frequency.

~

Finally, the likelihood function for wideband timing and noise analysis is \citep{AlamArzoumanian+2021,SusobhananVan_Haasteren2025}
\begin{equation}
    \ln L = 
    -\frac{1}{2}\textbf{y}^T \textbf{N}^{-1} \textbf{y} 
    - \frac{1}{2}\ln\det [2\pi\textbf{N}]\,,
    \label{eq:lnlike}
\end{equation}
where $\textbf{y}$ is a $2n_{\text{toa}}$-dimensional column vector containing the timing residuals and the DM residuals $(r_i,\delta_i)$ corresponding to each wideband measurement computed using equations \eqref{eq:delays}-\eqref{eq:dm-resids}, and 
$\textbf{N}$ is a $2n_{\text{toa}}\times 2n_{\text{toa}}$-dimensional TOA and DM covariance matrix, and $n_{\text{toa}}$ is the number of TOAs.

\updated{Some comments are in order regarding the structure of $\textbf{N}$.
In general, $\textbf{N}$ is a block-diagonal matrix containing the TOA and DM variances along the diagonal and off-diagonal elements representing the TOA-DM covariance at each epoch.
This TOA-DM covariance is a function of the frequency within the observing band at which the measurement is made, and can be removed in principle by adjusting the observing frequency.
\citet{PennucciDemorestRansom2014} estimates this zero-covariance frequency using a Hessian-based point estimate of the covariance matrix assuming a Gaussian distribution.
The \texttt{PulsePortraiture} package implements this frequency adjustment procedure, and assuming its accuracy, does not provide covariance estimates in its measurements.
However, \citet{PennucciDemorestRansom2014} themselves demonstrated that this frequency estimate is not always accurate, resulting in small non-zero covariances, especially when the signal-to-noise ratio is low.
Further, this procedure implicitly assumes that the noise present in the profile is uncorrelated across frequency channels.
This assumption is not valid when significant pulse jitter is present, since it is correlated across frequency channels \citep{KulkarniShannon+2024}.
Unfortunately, the currently available public wideband datasets, prepared using \texttt{PulsePortraiture}, do not provide any estimate of this covariance.
Therefore, despite the caveat above, we treat $\textbf{N}$ as a diagonal matrix in this paper.
This covariance will be investigated in detail in a future publication.
}

The first $n_{\text{toa}}$ diagonal elements of $\textbf{N}$ are $\varsigma_i^2$, and the rest are $\varepsilon_i^2$, where
\begin{subequations}
\begin{align}
    \varsigma_i^2 &= E^2 (\sigma_i^2 + Q^2)\,, \label{eq:scaled-toa-err}\\
    \varepsilon_i^2 &= \mathcal{E}^2 (\epsilon_i^2 + \mathcal{Q}^2)\label{eq:scaled-dm-err}\,.
\end{align}
\end{subequations}
$\sigma_i$ and $\epsilon_i$ are the measurement uncertainties in TOAs and DMs, respectively, and $\varsigma_i$ and $\varepsilon_i$ are known as the scaled TOA and DM uncertainties.
The quantities $E$ and $Q$ that modify the TOA uncertainties are known as EFACs (`error factors') and EQUADs (`errors added in quadrature'), respectively.
Similarly, $\mathcal{E}$ and $\mathcal{Q}$ are known as DMEFACs and DMEQUADs.
The EFACS, EQUADs, DMEFACs, and DMEQUADs characterize the uncorrelated part of the pulse jitter noise and imperfections in the TOA and DM measurement algorithms, leading to misestimation of uncertainties, and these depend on the observing system.

~

Let $\boldsymbol{a}$ represent the timing model parameters appearing in equations \eqref{eq:delays}-\eqref{eq:dm-resids} whose prior distributions are determined by the hyperparameters $\boldsymbol{A}$, and let $\boldsymbol{b}$ be the time-uncorrelated noise parameters appearing in equations \eqref{eq:scaled-toa-err}-\eqref{eq:scaled-dm-err}.
The log-posterior distribution of these parameters given a dataset $\mathfrak{D}$ and a timing and noise model $\mathfrak{M}$ can be written using Bayes theorem as
\begin{align}
    \ln P[\boldsymbol{a},\boldsymbol{b},\boldsymbol{A} | \mathfrak{D}, \mathfrak{M}] &= 
    \ln L[\mathfrak{D} | \boldsymbol{a},\boldsymbol{b}, \mathfrak{M}] 
     + \ln \Pi[\boldsymbol{a} | \boldsymbol{A},\mathfrak{M}]
      \nonumber\\
    &\quad + \ln \Pi[\boldsymbol{b} |  \mathfrak{M}] +\ln \Pi[\boldsymbol{A} | \mathfrak{M}]\nonumber\\
    &\quad - \ln Z[\mathfrak{D} | \mathfrak{M}]\,,
    \label{eq:lnpost}
\end{align}
where $\ln L[\mathfrak{D} | \boldsymbol{a},\boldsymbol{b}, \mathfrak{M}]$ is given by equation \eqref{eq:lnlike}, $\Pi$ represents the prior distributions, and the normalizing constant $Z$ is known as the Bayesian evidence.

~

The time-correlated stochastic processes appearing in the timing model, such as $\phi_\text{SN}$ and $\mathcal{D}_\text{DMN}$, are usually modeled as Gaussian processes in the Fourier domain \citep{LentatiAlexander+2014,van_HaasterenVallisneri2014a}.
Assuming their effect on the timing residuals to be small compared to the pulsar rotational period $F_0^{-1}$, we can express the timing residuals as a linear function in the noise amplitudes $\boldsymbol{\alpha}$ \citep{SusobhananVan_Haasteren2025}
\begin{equation}
 \textbf{y} = \textbf{y}' + \textbf{U}\boldsymbol{\alpha}\,,     
\end{equation}
where $\textbf{U}$ is a noise design matrix containing derivatives of $r_i$ and $\delta_i$ with respect to the amplitudes $\boldsymbol{\alpha}$.
Equation \eqref{eq:lnlike} can be written in this case as
\begin{align}
    \ln L[\boldsymbol{\theta},\boldsymbol{\alpha},\boldsymbol{b},\boldsymbol{A}|\mathfrak{D},\mathfrak{M}] &= 
    -\frac{1}{2}(\textbf{y} - \textbf{U}\boldsymbol{\alpha})^T \textbf{N}^{-1} (\textbf{y} - \textbf{U}\boldsymbol{\alpha}) \nonumber\\
    &\quad - \frac{1}{2}\ln\det [2\pi\textbf{N}]\,,
    \label{eq:lnlike-lin}
\end{align}
where we have separated $\boldsymbol{a}$ into the noise amplitudes $\boldsymbol{\alpha}$ and the remaining timing parameters $\boldsymbol{\theta}$.
Substituting this expression into equation \eqref{eq:lnpost} and imposing Gaussian priors $\Pi[\boldsymbol{\alpha}|\boldsymbol{A},\mathfrak{M}] = \mathcal{N}[0,\boldsymbol{\Phi}(\boldsymbol{A})]$, we can analytically marginalize the posterior distribution over $\boldsymbol{\alpha}$ \citep{SusobhananVan_Haasteren2025}.
This yields
\begin{align}
    \ln P[\boldsymbol{\theta},\boldsymbol{b},\boldsymbol{A} | \mathfrak{D}, \mathfrak{M}] &= 
    \ln \Lambda[\mathfrak{D} | \boldsymbol{\theta},\boldsymbol{b}, \mathfrak{M}] 
     + \ln \Pi[\boldsymbol{\theta} | \mathfrak{M}]\nonumber\\
      &\quad+ \ln \Pi[\boldsymbol{b} |  \mathfrak{M}]
    + \ln \Pi[\boldsymbol{A} | \mathfrak{M}]\nonumber\\
      &\quad    - \ln Z[\mathfrak{D} | \mathfrak{M}]\,,
    \label{eq:lnpost-marg}
\end{align}
with 
\begin{equation}
    \ln \Lambda[\boldsymbol{\theta},\boldsymbol{b},\boldsymbol{A}|\mathfrak{D},\mathfrak{M}] = 
    -\frac{1}{2}\textbf{y}^T \textbf{C}^{-1} \textbf{y} 
    - \frac{1}{2}\ln\det [2\pi\textbf{C}]\,,
    \label{eq:lnlike-marg}
\end{equation}
where $\textbf{C} = \textbf{N} + \textbf{U}\boldsymbol{\Phi}\textbf{U}^T$, $\boldsymbol{\Phi}$ being the prior covariance matrix imposed upon $\boldsymbol{\alpha}$.
The covariance matrix \textbf{C} can also include contributions corresponding to analytically marginalized timing parameters following \citet{van_HaasterenLevin2013} and \citet{SusobhananVan_Haasteren2025}, and this is useful for improving the efficiency of the sampler in exploring the parameter space by reducing its dimensionality.

~

The implementation of the wideband timing and noise model described above in \vela{} is summarized in the next section.

\section{Wideband Timing in \vela{}}
\label{sec:vela-impl}

\begin{figure*}[tt]
    \centering
    \includegraphics[width=1\linewidth]{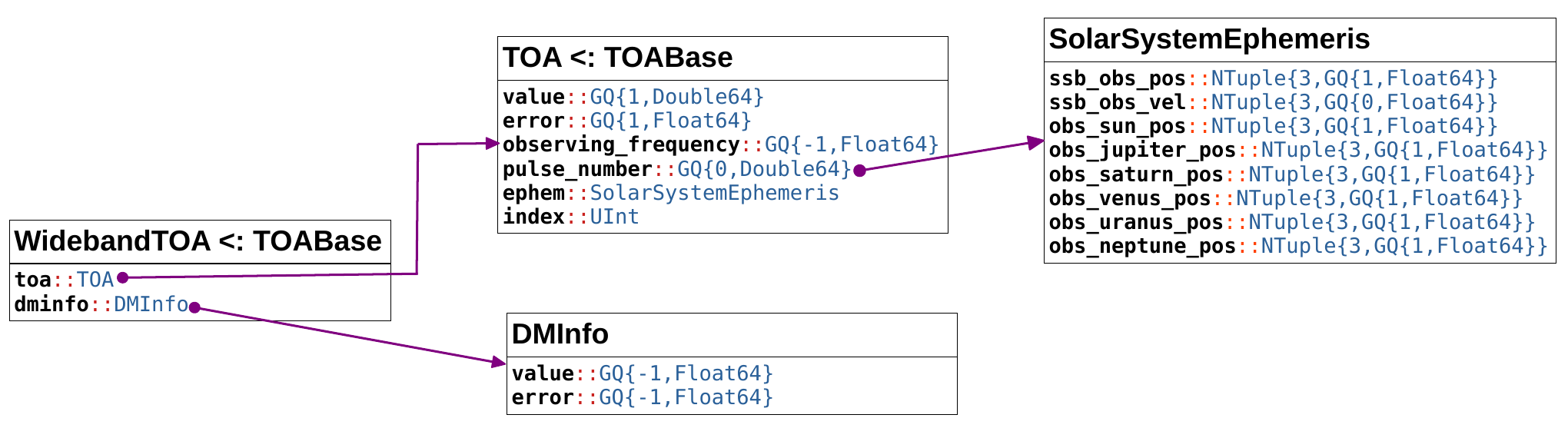}
    \caption{The structure of the \texttt{WidebandTOA} type. `\textless:' represents `subtype of'.
    The \texttt{GQ} type represents a quantity with dimensions $[\textsc{T}^d]$, and is described in \citet{Susobhanan2025}.}
    \label{fig:wtoa-type}
\end{figure*}

A wideband TOA and DM measurement is represented in \vela{} using the \texttt{WidebandTOA} type, containing a \texttt{TOA} object and a \texttt{DMInfo} object. 
The \texttt{TOA} type contains the clock-corrected TOA measurement, its uncertainty, observing frequency, solar system ephemeris, etc, and is also used for representing narrowband TOAs \citep{Susobhanan2025}.
The \texttt{DMInfo} type contains the DM measurement and uncertainty.
The structure of the \texttt{WidebandTOA} type is shown in Figure \ref{fig:wtoa-type}.

Similar to the narrowband case, the timing and noise model is represented in \vela{} using the \texttt{TimingModel} type.
A \texttt{TimingModel} contains an ordered collection of \texttt{Component}s which compute the delay, phase, DM, and/or uncertainty corrections applicable to each \texttt{WidebandTOA}.
In particular, the DM corrections appearing in equation \eqref{eq:dmmodel} are represented by subtypes of \texttt{DispersionComponent}, which in turn is a subtype of \texttt{Component}.
These types are also used for representing the timing and noise model in the narrowband case \citep{Susobhanan2025}, and we have extended the relevant functions such as \texttt{correct\_toa()} to handle wideband data.
See Table 2 and Figure 9 of \citet{Susobhanan2025} for a list of available \texttt{Components} and their type hierarchy.

The \texttt{TimingModel} also contains a \texttt{Kernel} which implements the log-likelihood computation given in equations \eqref{eq:lnlike} and \eqref{eq:lnlike-marg}.
In the wideband case, two types of  \texttt{Kernel} are available.
The \texttt{WhiteNoiseKernel} implements equation \eqref{eq:lnlike} and is applicable when the TOA and DM covariance matrix is diagonal.
The \texttt{WoorburyKernel} implements equation \eqref{eq:lnlike-marg} and is applicable to a more general class of covariance matrices which can be expressed in the reduced-rank form $\textbf{C} = \textbf{N} + \textbf{U}\boldsymbol{\Phi}\textbf{U}^T$ with a diagonal or block-diagonal $\textbf{N}$ matrix.\footnote{The \texttt{WoodburyKernel} was implemented after the publication of \citet{Susobhanan2025}. It also handles narrowband datasets following, e.g., \citet{LentatiAlexander+2014}. 
The name reflects the fact that the matrix inverse ($\textbf{C}^{-1}$) appearing in equation \eqref{eq:lnlike-marg} is evaluated using the Woodbury identity.}
The log-posterior evaluation for wideband datasets is parallelized using multi-threading similar to the narrowband case.
Further, the implementation of the prior distributions and the handling of model parameters is identical to the narrowband case described in \citet{Susobhanan2025}.

Partial analytic marginalization over a subset of timing parameters \citep{van_HaasterenLevin2013} for both narrowband and wideband datasets is implemented with the help of a special \texttt{Component} type named \texttt{MarginalizedTimingModel} that can only be used as part of a \texttt{WoodburyKernel}.
Only parameters whose effects on the timing residuals are exactly or approximately linear are allowed to be analytically marginalized. 
This currently includes the overall phase offset, rotational frequency, its derivatives, instrumental time jumps, DMX parameters, and wideband DM jumps.

Similar to the narrowband case, wideband timing and noise analysis is also accessible through the \texttt{pyvela} Python interface with identical syntax; see Figure \ref{algo:pyvela-emcee}.
In particular, the \texttt{SPNTA} type in \texttt{pyvela} also supports wideband data, and the \texttt{SPNTA.is\_wideband()} method can be used to check if a dataset is wideband. 
Additionally, we have also implemented a new command-line interface for \vela{}, and this is described in Appendix \ref{sec:pyvela-cli}.

\begin{figure*}
    \centering
    \includegraphics[width=1\linewidth]{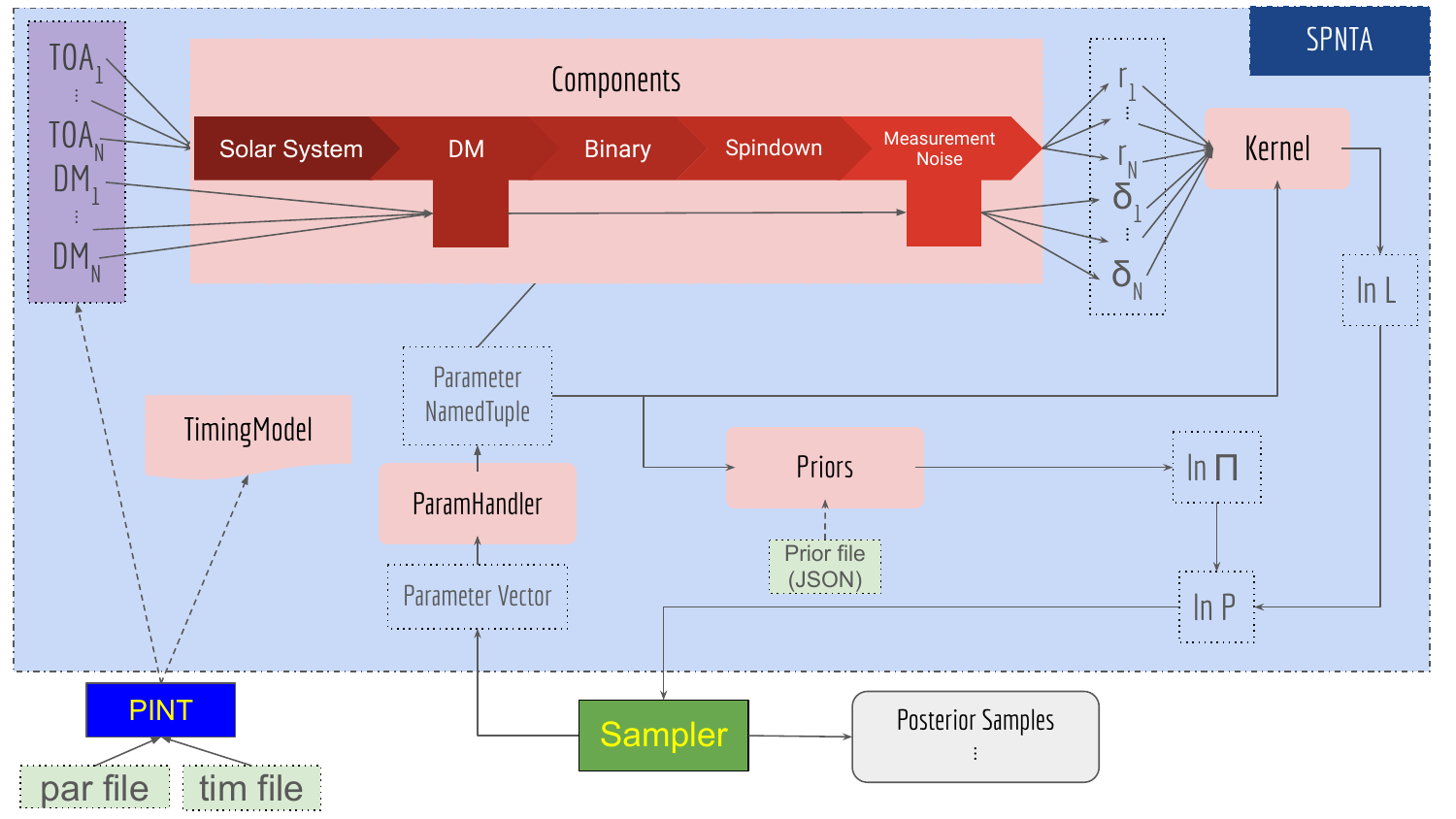}
    \caption{Schematic diagram summarizing \vela{}'s working in the wideband paradigm. 
    The \texttt{Component}s, the \texttt{Kernel}, the \texttt{ParamHandler}, and the \texttt{Prior}s are part of the \texttt{TimingModel} although they are shown separately for clarity.
    The `Measurement Noise' block includes EFACs, EQUADs, DMEFACs, and DMEQUADs, and acts on TOA and DM measurement uncertainties, which are not shown explicitly.
    Except for the \texttt{Component}s and \texttt{Kernel} that act on the DM measurements, all components of \vela{} behave identically to the narrowband case.}
    \label{fig:vela-schematic-wb}
\end{figure*}

The working of \vela{} in the wideband paradigm is illustrated in Figure \ref{fig:vela-schematic-wb}.
A Python script demonstrating the use of \vela{} is shown in Figure \ref{algo:pyvela-emcee}.
Figure \ref{fig:performance} shows the execution time of \vela{}'s log-likelihood computation as a function of the number of TOAs using simulated data.

\begin{figure*}
\begin{lstlisting}[language=Python]
from pyvela import SPNTA
import emcee
import numpy as np

spnta = SPNTA(
    parfile="J1234+5678.par",
    timfile="J1234+5678.tim", 
    custom_priors="J1234+5678_priors.json",
    cheat_prior_scale=100,
    analytic_marginalized_params=["PHOFF", "F0", "F1"],
)

nwalkers = spnta.ndim * 5
p0 = np.array([
    spnta.prior_transform(cube) 
    for cube in np.random.rand(nwalkers, spnta.ndim)
])

sampler = emcee.EnsembleSampler(
    nwalkers,
    spnta.ndim,
    spnta.lnpost,
)

sampler.run_mcmc(p0, 6000, progress=True)
samples_raw = sampler.get_chain(flat=True, discard=1000, thin=50)

samples = spnta.rescale_samples(samples_raw)
\end{lstlisting}
\caption{An example script demonstrating the usage of the \pyvela{} interface with \emcee{} \citep{Foreman-MackeyHogg+2013}. 
User-defined priors are read from a \texttt{JSON} file, and `cheat' priors are set based on the frequentist uncertainties scaled by \texttt{cheat\_prior\_scale}; see Appendix B of \citet{Susobhanan2025}.
The \texttt{analytic\_marginalized\_params} option specifies which parameters should be analytically marginalized.
The \texttt{emcee.EnsembleSampler} object is initialized with prior samples drawn with the help of \texttt{SPNTA.prior\_transform()}. 
The posterior samples are converted from \vela{}'s internal units to their common units using \texttt{SPNTA.rescale\_samples()}.
The high-level syntax shown here is identical for both narrowband and wideband datasets.}
\label{algo:pyvela-emcee}
\end{figure*}

\begin{figure}
    \centering
    \includegraphics[width=1\linewidth,trim={0.1cm 0.1cm 0.6cm 0.6cm},clip]{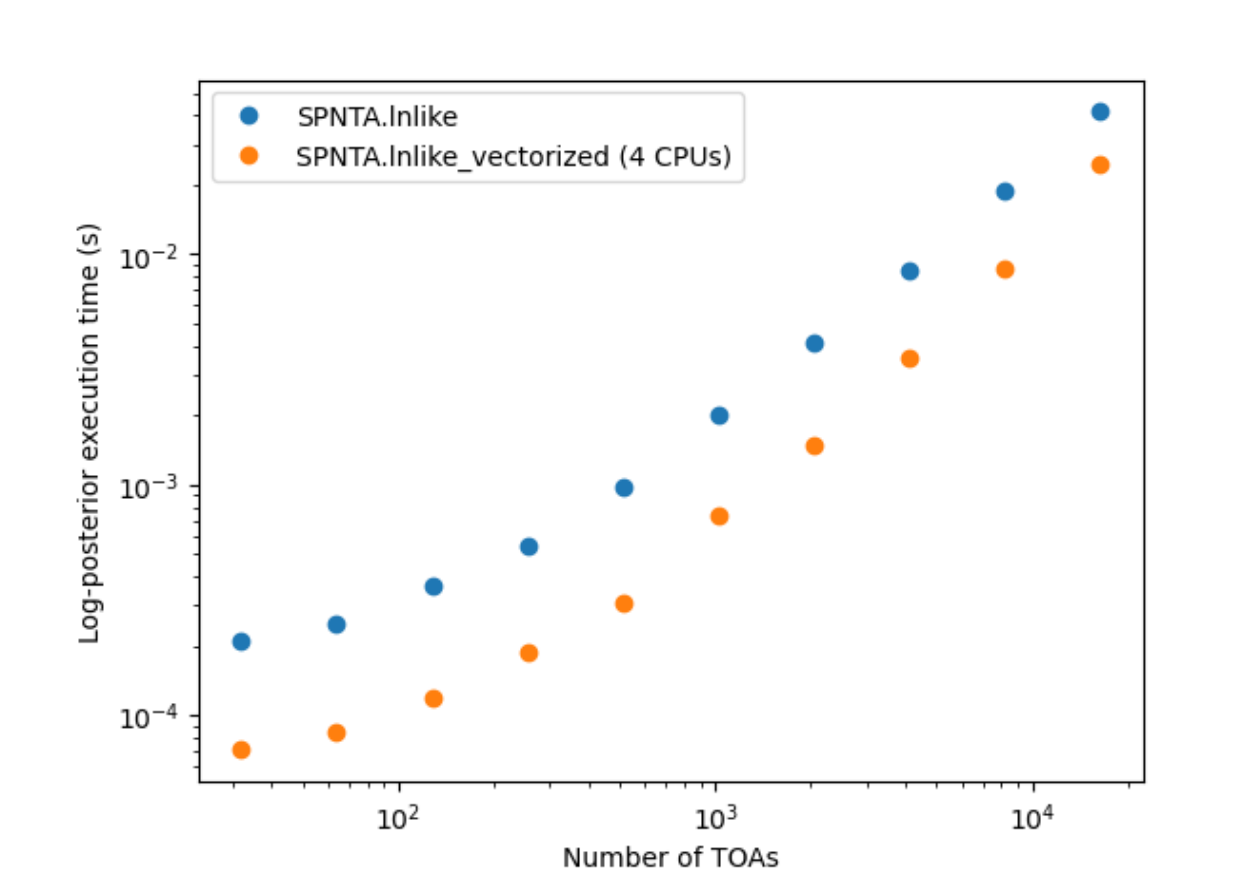}
    \caption{Execution time of \vela{}'s log-likelihood function as a function of number of TOAs for a simulated narrowband dataset. 
    The execution times for wideband datasets are similar.
    The timing model includes quadratic spin-down of the pulsar, spin noise modeled as a reduced-rank Gaussian process with 30 harmonics, the relativistic binary model of \citet{DamourDeruelle1986}, dispersion measure variations modeled using a quadratic polynomial and a reduced-rank Gaussian process with 30 harmonics,  solar system effects (including proper motion and parallax), and instrumental noise modeled using an EFAC, EQUAD, and a DMEFAC, and frequency-dependent profile evolution modeled using FD parameters \citep{ArzoumanianBrazier+2015}.
    The DM and spin noise amplitudes were analytically marginalized.
    The execution times were estimated on an AMD Ryzen 7 CPU with 16 GB of RAM. 
    The difference between \texttt{SPNTA.lnlike} and \texttt{SPNTA.lnlike\_vectorized} is that the former computes the likelihood function serially, whereas the latter computes the likelihood function in parallel using multiple threads for multiple points in the parameter space (100 points in this case).
    These numbers depend on the structure of the dataset, the complexity of the timing and noise model, and the executing machine, and therefore must be treated only as order-of-magnitude estimates. 
    }
    \label{fig:performance}
\end{figure}

\section{Application to PSR J1923+2515}
\label{sec:sims}
We now demonstrate the application of \vela{} using the NANOGrav 12.5-year wideband dataset of the isolated pulsar J1923+2515 \citep{AlamArzoumanian+2021}.
This dataset contains 119 wideband measurements (TOAs, DMs, and the corresponding uncertainties) taken at the Arecibo observatory between MJDs 55791--57927.
The TOA-DM covariances are assumed to be zero.
Two receivers, one centered around 1.4 GHz (`L-wide') and one centered around 430 MHz (`430'), and two backends (`ASP' and `PUPPI') were used to record the data.

The timing and noise model includes pulsar rotational frequency and its derivative, 
solar system delays, interstellar DM and its derivative, a time jump between the two receivers, wideband DM jumps corresponding to each receiver, EFACs, EQUADs, and DMEFACs corresponding to each receiver-backend combination, and reduced-rank Gaussian process models for DM noise and achromatic red noise with power law spectra.
\updated{Achromatic red noise includes the pulsar spin noise and possibly the effect of a gravitational wave background; these stochastic processes cannot be separated in a single-pulsar analysis.}
We use 16 linearly spaced harmonics for the DM noise and achromatic red noise with a fundamental frequency of 5.42 nHz \updated{corresponding to the total observing time span}.
Following \citet{van_HaasterenVallisneri2014a}, we also include 4 logarithmically spaced harmonics below the fundamental frequency to better model the lower frequency components of the DM and spin noise.
The solar system delays are computed using the DE440 solar system ephemeris \citep{ParkFolkner+2021}, and the clock corrections are computed using the BIPM2021 realization of the TT timescale.

Following \citet{SusobhananVan_Haasteren2025}, we analytically marginalize the DM noise and achromatic red noise amplitudes as well as the overall phase offset, rotational frequency, its derivative, instrumental time jumps, and wideband DM jumps \updated{assuming improper Gaussian priors with infinite width} to reduce the dimensionality of the parameter space.
We draw samples from the partially marginalized posterior distribution using the \emcee{} package, which implements the affine-invariant ensemble sampler algorithm \citep{Foreman-MackeyHogg+2013}.
The Python script used to perform this analysis is similar to the one shown in Figure \ref{algo:pyvela-emcee}. 

\updated{The prior distributions used in this analysis are listed in Table \ref{tab:J1923+2515-params}.
The prior distribution on parallax is based on a distance estimate obtained from the DM using the YMW16 electron density model \citep{YaoManchesterWang2017}.
This pulsar is at a DM of around 18.86 pc/cm$^3$, and this corresponds to a distance of $\sim$1200 kpc (parallax of $\sim$0.83 mas) at its sky location.
Unfortunately, DM-based distance estimates are known to suffer from large systematic errors.
Therefore, we set the parallax prior distribution to be a Gaussian with a mean of 0.83 mas and a 50\% standard deviation, truncated at 0.
We use log-normal priors for EFACs and DMEFACs so that they remain positive, but favor values close to 1.
Log-uniform priors are used for EQUADs, and uniform priors are used for achromatic red noise and DM noise log-amplitudes and spectral indices.
The data strongly constrain the sky coordinates, proper motion, DM, and DM derivative, and we use uniform `cheat' priors for these parameters centered around the frequentist estimate, with a width equal to 100 times the frequentist uncertainty (see Appendix B of \citet{Susobhanan2025} for a detailed discussion on cheat priors).
We have checked that increasing the width of the `cheat' priors has no discernible impact on the posterior distribution.
}

\begin{figure*}
    \centering
    \includegraphics[width=1\linewidth]{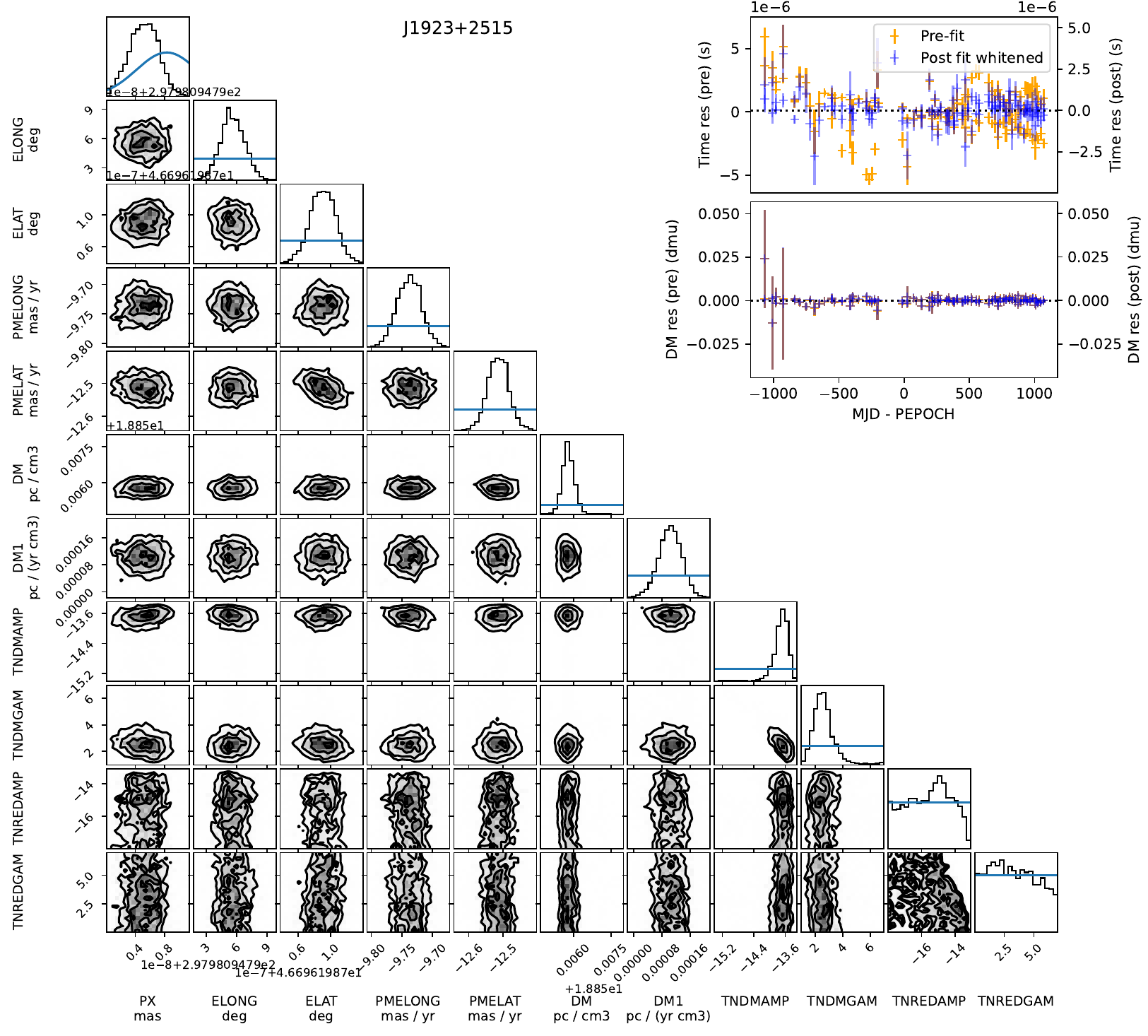}
    \caption{The posterior distribution for timing and noise model parameters for PSR J1923+2515 as described in Section \ref{sec:sims}.
    Blue curves plotted with marginalized univariate posterior distributions represent the corresponding prior distribution. 
    The parameters and prior distributions shown in this figure are described in Table \ref{tab:J1923+2515-params}. 
    EFACs, EQUADs, and DMEFACs, as well as the analytically marginalized parameters, are not plotted.
    The time and DM residuals obtained using the median sample are plotted in the inset, along with their pre-fit counterparts.
    PEPOCH is the spin frequency epoch, and its value in MJD is 56859.}
    \label{fig:J1923+2515-results}
\end{figure*}

\begin{figure}
    \centering
    \includegraphics[width=1\linewidth]{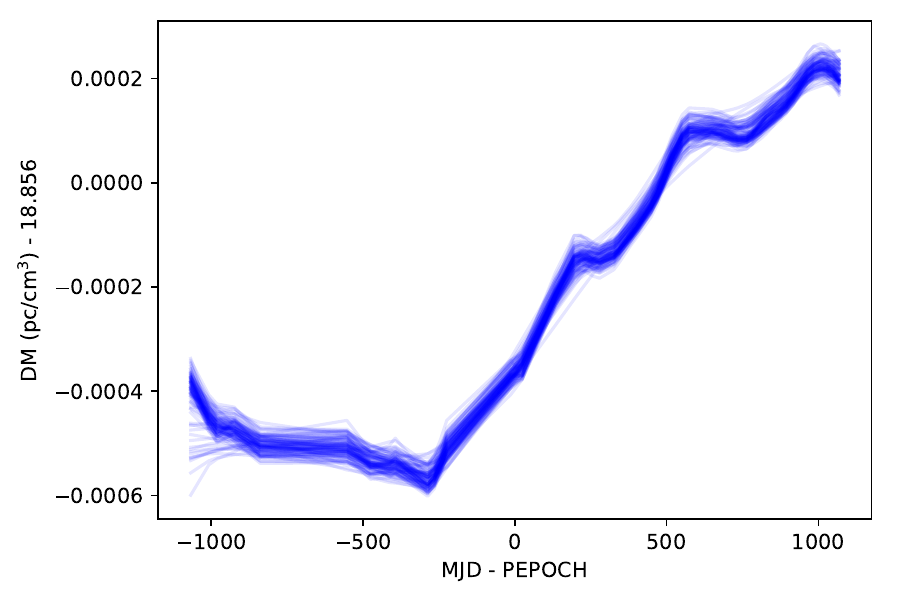}
    \caption{Hundred random DM time series realizations of PSR J1923+2515.
    These were generated using parameters drawn from the posterior distribution shown in Figure \ref{fig:J1923+2515-results}. PEPOCH is the spin frequency epoch, and its value in MJD is 56859.}
    \label{fig:model-dm}
\end{figure}

The posterior distribution and the post-fit residuals obtained from this analysis are plotted in Figure \ref{fig:J1923+2515-results}, and the measurement/constraint for each parameter obtained therefrom is listed in Table \ref{tab:J1923+2515-params}.
\updated{The prior distributions for the different parameters are also plotted in Figure \ref{fig:J1923+2515-results} for comparison.}
Proper motion is measured, \updated{but parallax shows a marginal detection that is consistent with 0 within the 3$\sigma$ credible interval}.
Achromatic red noise is not detected, and the posterior distribution shows a 95.45\% upper limit of -13.7 for its log-amplitude.
This upper limit is consistent with the amplitude of the recently found common achromatic signal, possibly due to a gravitational wave background, discussed in \citet{AgazieAntoniadis+2024}. 
DM noise is detected, and a hundred random DM time series realizations generated using parameters drawn from the posterior distribution are plotted in Figure \ref{fig:model-dm}.

\begin{table*}
\begin{tabular}{clcc}
\hline
 &                       &  & \textbf{Measurement/} \\
\textbf{Parameter} &               \textbf{Description \& Unit}        & \textbf{Prior distribution} & \textbf{Constraint} \\
\hline
ELONG               & Ecliptic longitude (deg)                                          &                      & $297.98094796(1)$       \\
ELAT              & Ecliptic latitude (deg)                         &   & 46.69619879(2)                                              \\
PMELONG & Proper motion in ecliptic longitude (mas/yr) & & -9.74(2) \\
PMELAT & Proper motion in ecliptic latitude (mas/yr) & & -12.51(3) \\
PX & Parallax (mas) & Trunc[Normal[0.83, 0.415], 0, $\infty$] & 0.5(2) \\
DM                 & Dispersion measure (pc/cm$^3$)                                              &       & 18.8558(3)                      \\
DM1                & Dispersion measure derivative (pc/cm$^3$/yr)                     &     & 0.00010(4)                        \\
EFAC              & EFAC for each receiver-backend combination                                             & LogNormal{[}0.0, 0.25{]}   \\
EQUAD              & EQUAD for each receiver-backend combination                                            & LogUniform{[}10$^{-3}$, 10$^2${]}   \\
DMEFAC            & DMEFAC form each receiver-backend combination                                          & LogNormal{[}0.0, 0.25{]}   \\
TNREDAMP            & Achromatic red noise log-amplitude                                          & Uniform[-18, -9]  & -13.7$\dag$ \\
TNREDGAM            & Achromatic red noise spectral index                                            & Uniform[0, 7]  & \textit{Not measured} \\
TNDMAMP            & DM noise log-amplitude                                          & Uniform[-18, -9] &-13.7(2)  \\
TNDMGAM            & DM noise spectral index                                            & Uniform[0, 7] & 2.7(8)  \\
\hline
\end{tabular}
\caption{The timing and noise model parameters estimated in Section \ref{sec:sims}. 
The prior distribution and the measurement/constraint for each parameter derived from the posterior samples are listed.
Analytically marginalized parameters (overall phase offset, rotational frequency and its derivative, instrumental time jumps, and instrumental DM jumps) are not listed; irregular Gaussian priors with infinite variance are assumed for these parameters.
\updated{`Trunc' indicates a truncated distribution.}
The DM derivative is defined at the epoch MJD 56859.
`Cheat' priors, namely uniform priors centered at the frequentist estimate (obtained using \pint{}) with a width equal to 100 times the frequentist uncertainty, are used for some parameters, and such priors are not listed in the table (see Appendix B of \citet{Susobhanan2025} for a detailed explanation).
We have checked that increasing the `cheat' prior widths does not appreciably change the posterior distribution. 
The marginalized posterior distribution for the spin noise spectral index (TNREDGAM) is consistent with its prior distribution, and this is indicated as `Not measured' in the table.
The posterior distribution constrains the achromatic red noise log-amplitude from above, and its 95.45\% upper limit is indicated with `$\dag$'.
The measurements/constraints for the white noise parameters EFAC, EQUAD, and DMEFAC are not listed since they are not astrophysically relevant.}
\label{tab:J1923+2515-params}
\end{table*}

\section{Summary}
\label{sec:summary}

We implemented support for wideband timing in the \vela{} Bayesian pulsar timing and noise analysis package.
This enables \vela{} to compute the log-likelihood, log-prior, prior transform, and log-posterior functions for wideband timing datasets using the full non-linear timing model, optionally marginalizing over the noise amplitudes as well as a subset of timing parameters whose effect on the residuals is approximately linear. 
\vela{} is the first package to have this functionality.
The log-likelihood and log-posterior computations can be parallelized across multiple CPU cores using multi-threading.

We extended the \pyvela{} Python interface of \vela{} such that it provides an interface for wideband datasets that is \textit{identical} to the one for narrowband datasets.
This can be used alongside any Markov Chain Monte Carlo \citep{Diaconis2009} or nested sampling \citep{AshtonBernstein+2022} library to perform Bayesian inference.
We have also developed the \pyvela{} command-line interface, which can be used to run quick analyses when fine-tuning of the timing and noise model and the sampler is not necessary, combining \vela{} with the \emcee{} sampling package.
We demonstrated the application of \vela{} on the NANOGrav 12.5-year wideband dataset of PSR J1923+2515.
This development should enable, for the first time, Bayesian single-pulsar noise and timing analysis to be performed on wideband datasets, including PTA datasets.

The implementation of wideband timing in \vela{} is summarized below.
\begin{enumerate}
    \item A clock-corrected wideband TOA and DM measurement is represented using the \texttt{WidebandTOA} type.
    \item A timing and noise model is represented using the \texttt{TimingModel} type. Its structure remains the same as in the narrowband case, but we have extended the various functions operating on \texttt{TimingModel} and its constituents to handle \texttt{WidebandTOA}s.
    \item The \pyvela{} Python interface retains the same interface as in narrowband timing, but now also handles wideband datasets.
\end{enumerate}

The planned future developments in \vela{} include the implementation of photon-domain timing for high-energy observations \citep{PletschClark+2015}, the implementation of a sampler optimized for single-pulsar noise \& timing analysis \citep[e.g., following ][]{LaalLamb+2023}, reconstruction of cross-pulsar correlations from SPNTA results \citep{ValtolinaVan_Haasteren2024}, and the investigation of pulse jitter-induced correlations in wideband measurements. 
A more general framework for linear fitting of wideband datasets based on this work and \citet{SusobhananVan_Haasteren2025} is being implemented in \pint{}.
Further, a non-linear fitting algorithm for \pint{} using \vela{} is also under development.

\begin{acknowledgments}
I thank David Kaplan for providing valuable feedback on the manuscript and for contributing to the development and testing of \vela{}.
I thank Rutger van Haasteren for fruitful discussions, Chiara Mingarelli for valuable comments on this manuscript, and Paul Ray for valuable feedback on the \vela{} package.
\end{acknowledgments}

\section*{Software}
\pint{} \citep{LuoRansom+2021, SusobhananKaplan+2024}, 
\emcee{} \citep{Foreman-MackeyHogg+2013}, 
\texttt{pygedm} \citep{PriceFlynnDeller+2021},
\texttt{numpy} \citep{HarrisMillman+2020}, 
\astropy{} \citep{RobitailleTollerud+2013}, 
\texttt{matplotlib} \citep{Hunter2007}, 
\texttt{corner} \citep{Foreman-Mackey2016}, 
\texttt{DoubleFloats.jl} \citep{Sarnoff2022}, \texttt{Distributions.jl} \citep{BesançonPapamarkou+2021}, \texttt{JuliaCall/PythonCall.jl} \citep{Rowley2022},
\texttt{JLSO.jl} \citep{FinneganWhite2022},
\texttt{git}\footnote{\url{https://git-scm.com/}}.

\section*{Data Availability}

The NANOGrav 12.5-year dataset is available at \url{https://nanograv.org/science/data/125-year-pulsar-timing-array-data-release}.

\appendix

\section{The \texttt{pyvela} command-line interface}
\label{sec:pyvela-cli}
The \texttt{pyvela} script (not to be confused with the \texttt{pyvela} Python interface) provides a convenient way to run a Bayesian pulsar timing and noise analysis from the command line using \texttt{Vela.jl} in combination with the \texttt{emcee} sampler.
It sacrifices flexibility and fine-tuning capabilities in favor of convenience.

This script takes a pair of \texttt{par} and \texttt{tim} files, a prior file in the \texttt{JSON} format, and an output directory as input, and saves the posterior samples and related metadata such as parameter names, units, etc, into the output directory.
An example invocation of the \pyvela{} script is given below:
\begin{lstlisting}[language=bash]
pyvela J1234-5678.par J1234-5678.tim 
  -o J1234-5678_out
  -A PHOFF F0 F1 -C 200 
  -P J1234-5678_priors.json 
  -N 6000 -b 1500 -t 40
\end{lstlisting}
Here, the first two positional arguments are the \texttt{par} and \texttt{tim} files, the \texttt{-o} option specifies the output directory, the \texttt{-A} option enables partial analytic marginalization over a subset of timing parameters, the \texttt{-P} option provides the prior file, \texttt{-C} option specifies the factor with which to multiply the frequentist uncertainties to obtain the `cheat' priors of parameters whose priors are otherwise unspecified, \texttt{-N} option specifies the number of iterations for the sampler, \texttt{-b} option specifies the burn-in length, and \texttt{-t} option specifies the thinning factor for the posterior samples.

Other scripts included with \vela{} include \texttt{pyvela-plot}, which creates a plot similar to Figure \ref{fig:J1923+2515-results} given \pyvela{} output, \texttt{pyvela-rethin}, which modifies the burn-in length and thinning factor of the posterior samples, and \texttt{pyvela-jlso}, which saves the data, model, and prior distributions into a \texttt{JLSO} file \citep{FinneganWhite2022} for faster reading.

\vfill

\bibliography{vela-wideband}

\end{document}


\section*{\centering Glossary of symbols}

\begin{longtable}{|c|l|}
\hline 
\textbf{Symbol} & \textbf{Description}\tabularnewline
\hline 
\hline 
\multicolumn{2}{|c|}{\emph{{Data}}}\tabularnewline
\hline 
$\mathfrak{D}$ & Dataset as a whole\tabularnewline
$t_{\text{arr}}$ & Measured TOA\tabularnewline
$d$ & Measured DM\tabularnewline
$\sigma$ & Measured TOA uncertainty\tabularnewline
$\epsilon$ & Measured DM uncertainty\tabularnewline
$n_{\text{toa}}$ & Number of TOAs\tabularnewline
\hline 
\multicolumn{2}{|c|}{\emph{{Timing and noise model}}}\tabularnewline
\hline 
$\mathfrak{M}$ & Timing and noise model as a whole\tabularnewline
$t_{\text{em}}$ & Estimated time of emission\tabularnewline
$\Delta_{\text{B}}$ & Binary delays\tabularnewline
$\Delta_{\text{DM}}$ & Dispersion delays\tabularnewline
$\Delta_{\text{GW}}$ & Gravitational wave-induced delays\tabularnewline
$\Delta_{\odot}$ & Solar system delays\tabularnewline
$\Delta_{\text{clock}}$ & Clock corrections\tabularnewline
$\Delta_{\text{jump}}$ & Instrumental jumps\tabularnewline
$\mathcal{N}_{\text{R}}$ & Radiometer noise affecting TOAs\tabularnewline
$\mathcal{N}_{\text{jitter}}$ & Jitter noise affecting TOAs\tabularnewline
$\mathcal{D}$ & Total astrophysical DM\tabularnewline
$\mathcal{D}'_{\text{jump}}$ & Instrumental wideband DM jumps\tabularnewline
$\mathcal{M}_{\text{R}}$ & Radiometer noise affecting DMs\tabularnewline
$\mathcal{M}_{\text{jitter}}$ & Jitter noise affecting DMs\tabularnewline
$D_{n}$ & Taylor series coefficients of DM variations\tabularnewline
$n_{D}$ & Number of Taylor series coefficients of DM variations\tabularnewline
$t_{0}$ & Fiducial epoch\tabularnewline
$D_{\text{DMN}}$ & Interstellar DM variations\tabularnewline
$D_{\text{SW}}$ & Solar wind DM\tabularnewline
$\phi$ & Estimated pulse phase\tabularnewline
$\phi_{0}$ & Overall phase offset\tabularnewline
$F_{n}$ & Taylor series coefficients of rotational frequency\tabularnewline
$n_{F}$ & Number of Taylor series coefficients of rotational frequency\tabularnewline
$\phi_{\text{glitch}}$ & Glitch phase\tabularnewline
$\phi_{\text{SN}}$ & Spin noise phase\tabularnewline
$\bar{F}$ & Topocentric pulsar frequency\tabularnewline
$E$ & EFAC\tabularnewline
$Q$ & EQUAD\tabularnewline
$\mathcal{E}$ & DMEFAC\tabularnewline
$\mathcal{Q}$ & DMEQUAD\tabularnewline
$\boldsymbol{a}$ & Collection of all timing parameters\tabularnewline
$\boldsymbol{b}$ & Collection of all white noise parameters\tabularnewline
\textbf{$\boldsymbol{\alpha}$} & Collection of correlated noise amplitudes\tabularnewline
\textbf{$\boldsymbol{\theta}$} & Collection of all timing parameters except \textbf{$\boldsymbol{\alpha}$}\tabularnewline
$\boldsymbol{\Phi}$ & Prior covariance matrix on \textbf{$\boldsymbol{\alpha}$}\tabularnewline
$\boldsymbol{A}$ & Hyperparameters that determine $\boldsymbol{\Phi}$\tabularnewline
\textbf{U} & Correlated noise basis matrix\tabularnewline
\hline 
\multicolumn{2}{|c|}{\emph{{Residuals and data covariance matrices}}}\tabularnewline
\hline 
$r$ & Time residual\tabularnewline
$\delta$ & DM residual\tabularnewline
$\varsigma$ & Scaled TOA uncertainty\tabularnewline
$\varepsilon$ & Scaled DM uncertainty\tabularnewline
\textbf{y} & Vector containing time and DM residuals\tabularnewline
\textbf{y'} & \textbf{y} vector evaluated using the best-fit parameters\tabularnewline
\textbf{N} & Diagonal covariance matrix containing scaled TOA and DM uncertainties\tabularnewline
\textbf{C}  & Non-diagonal covariance matrix appearing in $\Lambda$ (see below)\tabularnewline
\hline 
\multicolumn{2}{|c|}{\emph{{Probabilities and Probability distributions}}}\tabularnewline
\hline 
$L$ & Likelihood function\tabularnewline
$\Pi$ & Prior distribution\tabularnewline
$P$ & Posterior distribution\tabularnewline
$Z$ & Bayesian evidence\tabularnewline
$\Lambda$ & Partially marginalized likelihood function \tabularnewline
\hline 
\end{longtable}